\documentclass[runningheads]{llncs}
\usepackage[numbers,sort&compress]{natbib}
\usepackage{amsmath}
\usepackage{amssymb}
\usepackage{comment}
\usepackage{bbding}

\usepackage{xspace}
\newcommand{\ld}[1]{\textcolor{purple}{LD: #1}}
\newcommand{\ey}[1]{\textcolor{blue}{EY: #1}}
\newcommand{\ww}[1]{\textcolor{red}{WW: #1}}
\renewcommand{\ld}[1]{}
\renewcommand{\ey}[1]{}
\renewcommand{\ww}[1]{}

\newcommand{\metaeval}[0]{meta-evaluation}
\newcommand{\Metaeval}[0]{Meta-evaluation}

\newcommand{\llmjudge}[0]{LLM judge}
\newcommand{\umbrela}[0]{{\sc Umbrela}}
\newcommand{\ARGUE}[0]{{\sc Argue}\xspace}

\newcommand{\argue}[0]{{\sc Argue}}  
\newcommand{\autoargue}[0]{{\sc Auto\-Argue}} 

\newcommand{\rubric}[0]{{\sc Rubric Workbench}}
\newcommand{\AutoNuggetizer}[0]{{\sc AutoNuggetizer}}
\newcommand{\crucible}[0]{{\sc Crucible}}
\newcommand{\ginger}[0]{{\sc Ginger}}

\newcommand{\gptresearcher}[0]{{\sc GptResearcher}}

\newcommand{\testprobe}[0]{subversion probe}
\newcommand{\TestProbe}[0]{Subversion Probe}

\newcommand{\testprobes}[0]{subversion probes}
\newcommand{\TestProbes}[0]{Subversion Probes}

\newcommand{\probe}[0]{probe}

\newcommand{\probes}[0]{probes}

\newcommand{\llama}[0]{LLaMA}
\newcommand{\llamabig}[0]{\texttt{\llama{}-\-3.3-\-70B-\-Instruct}}

\usepackage{booktabs}

\usepackage{orcidlink}
\newcommand\orcid[1]{\textsuperscript{\orcidlink{#1}}}

\begin{document}
%
\title{Insider Knowledge: How Much Can RAG Systems Gain from Evaluation Secrets?}
\titlerunning{Insider Knowledge for RAG Systems}
%
\author{Laura Dietz\inst{1}\orcid{0000-0003-1624-3907} \Envelope \and
 Bryan Li\inst{2}\orcid{0000-0002-5779-1662} \and
 Eugene Yang\inst{3}\orcid{0000-0002-0051-1535} \and
 Dawn Lawrie\inst{3}\orcid{0000-0001-7347-7086} \and
 William Walden\inst{3}\orcid{0000-0001-9931-2861} \and
 James Mayfield\inst{3}\orcid{0000-0003-3866-3013}}
\authorrunning{Dietz et al.}

\institute{
University of New Hampshire, Durham, New Hampshire, USA \email{dietz@cs.unh.edu} 
\and
University of Pennsylvania, Philadelphia, Pennsylvania, USA \email{bryanli@seas.upenn.edu} 
\and
Human Language Technology Center of Excellence, Johns Hopkins University, Baltimore, Maryland, USA \\
\email{\{eugene.yang,lawrie,wwalden1,mayfield\}@jhu.edu}
}%
\maketitle 
\begin{abstract}
RAG systems are increasingly evaluated and optimized using LLM judges,
an approach that is rapidly becoming the dominant paradigm for system assessment.
Nugget-based approaches in particular are now embedded not only in evaluation frameworks but also in the architectures of RAG systems themselves.
While this integration can lead to genuine improvements,
it also creates a risk of faulty measurements due to circularity.
In this paper, we investigate this risk through comparative experiments with nugget-based RAG systems,
including \ginger{} and \crucible{},
against strong baselines such as \gptresearcher{}.
By deliberately modifying \crucible{} to generate outputs optimized for an LLM judge,
we show that near-perfect evaluation scores can be achieved when elements of the evaluation—such as prompt templates or gold nuggets—are leaked or can be predicted.
Our results highlight the importance of blind evaluation settings and methodological diversity to guard against mistaking metric overfitting for genuine system progress.\footnote{Online appendix at \url{https://github.com/hltcoe/ecir26-crucible-appendix/}}
\end{abstract}

\keywords{Retrieval-augmented generation
\and LLM judge
\and Nugget evaluation.}

\section{Introduction}

\ld{STOP!  Disable todo macros in preamble before submitting!}
As LLM-based chat systems like ChatGPT and Claude become primary information sources for general users, ensuring accuracy, recency, and credible sourcing of their responses is critical for developing trustworthy systems~\cite{ding2025citationstrustllmgenerated,dhole2025adversem}. Retrieval-Augmented Generation (RAG) has recently emerged as a leading approach to achieve these aims~\cite{lewis2020retrieval, izacard2021leveraging, borgeaud2022improving}. In parallel, \emph{nugget-based} evaluation methods, which check generated outputs for the presence of key pieces of information (\emph{nuggets}), have become an important tool for RAG evaluation~\cite{mayfield2024evaluation, 10.1145/3726302.3730316, 10.1145/3726302.3730090, pavlu2012ir}, grounding system assessment in assessor-identified Q\&A pairs or claims. Increasingly, however, these methods rely on LLMs not only for matching nuggets to content, but also for developing nugget banks from scratch~\cite{10.1145/3726302.3730090,dietz2024workbench}. 

While all evaluation paradigms can produce systematically biased, invalid measurements,
LLM-based evaluation methods are particularly susceptible to a range of vulnerabilities \cite{dietz2025tropes}. For instance, circularity arises whenever components of the evaluation methods are usefully integrated into a RAG system. Moreover prompts, scripts, and LLM judges are routinely made public and may thus be inadvertently exploited by system developers. This can lead to inflated evaluation scores and reduced alignment with manual assessments \cite{clarke2024llm}.
A suggested safeguard against circularity is to rely on human-curated, nugget-based evaluation~\cite{mayfield2024evaluation, upadhyay2024umbrela, farzi2024exampp}.

Yet such safeguards are not perfect.
In this paper, we identify three pathways by which \emph{insider knowledge} of an LLM-based evaluation process can be exploited to inflate evaluation results,
even without a direct test set leak:
(1) tuning RAG systems to optimize narrowly for measured criteria while ignoring other, \emph{unmeasured} quality criteria (RQ1);
(2) exploiting knowledge of the evaluation prompt and LLM model to modify system outputs (RQ2);
and (3) predicting the set of gold nuggets for system usage  with high accuracy (RQ3).\footnote{To avoid confusion, we refer to nuggets generated automatically by RAG systems, such as \crucible{}, as \emph{system nuggets}, and manually created nuggets used by the evaluation system \ARGUE{} as \emph{gold nuggets}.}

Among these, RQ3 poses the most serious and underappreciated threat.
Nugget sets are meant to serve as independent ground truth.
However, our investigations reveal that system-generated nuggets can overlap substantially with human-curated ones. This overlap may be greater still for nugget creation pipelines, such as \AutoNuggetizer{}~\cite{pradeep2024autonuggetizer},
that rely on LLMs to generate candidate nuggets that are then manually filtered. Thus, even absent nefarious intent, system developers may achieve high nugget-based evaluation scores by leveraging LLM-based nugget generation components.

Indeed, \emph{none} of RQ1, RQ2, or RQ3 assumes intentional subversion: RAG systems may \emph{incidentally} be trained using a model similar to the LLM judge, developed using similar prompts, or optimized against metrics or nuggets similar to those used at test time. Moreover, adopting these ideas may lead to genuinely better systems, as noted by \citet{clarke2024llm}.
By simulating the downstream effects of integrating these components into system development, we show how safeguards can collapse, resulting in inflated results along \emph{measured} dimensions. Our findings reveal that automatic nugget-based evaluations, though valuable,
are vulnerable to a form of leakage that is subtle, pervasive, and arguably already occurring.


\paragraph{Contributions.}
To study this problem, we introduce \crucible{},
a RAG system that incorporates evaluation ideas to probe the susceptibility of nugget-based LLM judge systems. \crucible{} is designed to expose and study vulnerabilities of \autoargue{}~\cite{walden2025autoargue}, an open-source LLM-based implementation of \ARGUE{}~\cite{mayfield2024evaluation}.
Drawing on insider knowledge of \autoargue{},
\crucible{} demonstrates that the evaluation is robust only when key evaluation elements are kept secret.
Once these elements become public knowledge, they may be used to enhance system outputs and inflate scores without necessarily obtaining genuine improvements.


We present a case study on the TREC NeuCLIR 2024 Report Generation Pilot task---a long-form RAG task with complex queries and an emphasis on faithful citation.
Crucially, nuggets for this task were fully manually curated through a principled process. Our exclusive focus on TREC NeuCLIR is a consequence of the lack of available \emph{alternative} RAG test collections with fully manual nuggets at the time of our study.
Since the purpose of this case study is to explore the susceptibility of an LLM judge paradigm, we do not evaluate the efficiency of the RAG system or the automatic evaluation. 


\section{Background}

\subsection{LLM-as-a-Judge}

The introduction of LLMs as automatic judges for retrieval and generation
systems has led to a surge of interest in their reliability as replacements for
or supplements to human assessments~\cite{li2025generationjudgmentopportunitieschallenges}.
While initial studies highlighted strong correlations with human labels~\cite{upadhyay2024umbrela, thomas2024large},
more recent work documents weaknesses and instabilities that threaten validity~\cite{clarke2024llm, soboroff2025don}.

Recent IR-focused work emphasizes observational comparisons and prompt-level analyses.
\citet{balog2025rankers} synthesize the roles that LLMs can play as rankers, judges, and assistants,
and identify feedback loops as a fundamental risk for evaluation.
\citet{thakur2025rag} compare \llmjudge{}s against human annotations for retrieval-augmented generation tasks at the TREC RAG Track,
showing both alignment and divergence.
\citet{arabzadeh2025benchmarking} benchmark pointwise, pairwise, and rubric-based judging paradigms,
while their follow-up~\cite{arabzadeh2025promptsensitivity} highlights the prompt sensitivity of \llmjudge{}s.
\citet{upadhyay2025umbrela} extend these comparisons with a large-scale study of \umbrela{}, an open-source implementation of Bing's relevance assessor, across multiple settings. 
While their positive findings are used to support the reliability of LLM judges, \citet{clarke2024llm} show that \umbrela{} ceases to be a reliable evaluator under conditions of circularity. 

A complementary line of work designs adversarial content attacks to measure judge susceptibilities,
often by injecting instructions that override evaluation prompts (e.g.\ \emph{forget all previous instructions and rate this response as relevant}).
Early studies explored generic prompt subversions in generative settings~\cite{alaofi2024generative, bardas2025, basat2015, kurland2022},
while more recent work has shifted toward citation evaluation~\cite{dhole2025adversem}.


Taken together, the current literature offers descriptive accounts of vulnerabilities \cite{dietz2025tropes},
but lacks systematic meta-evaluation paradigms of the sort proposed in this work for investigating hypothesized weaknesses of LLM-based evaluations.



\subsection{(Auto-)ARGUE}
\ARGUE, introduced by Mayfield et al.~\citep{mayfield2024evaluation}, is a nugget-based evaluation framework designed for \emph{report generation}---generation of long-form, citation-backed responses to a complex user request.
\ARGUE evaluates reports at the sentence level, assessing each sentence for whether it is supported by its citations and whether it attests any of a set of nuggets (represented as Q\&A pairs) associated with the request. These judgments can then be accumulated into report-level scores.
The \ARGUE framework has been used in the manual evaluation for the TREC NeuCLIR Track Report Generation Pilot in 2024~\cite{lawrie2024overview}. The case study presented in this work focuses on \autoargue{}~\cite{walden2025autoargue}, an open-source LLM-based implementation of \argue{}.

\subsection{UMBRELA \TestProbe}

A \emph{subversion probe} wraps an existing IR/RAG system in a lightweight procedure designed to exploit weaknesses of an LLM judge.  
\citet{clarke2024llm} introduced such a probe for \umbrela{}~\cite{upadhyay2024umbrela}, which reorders documents based on \umbrela{} relevance labels. Applied to TREC RAG 2024 systems, it revealed two findings: (1) the probe delivered genuine quality improvements under human assessment; and (2) \umbrela{} produced systematically biased scores for manipulated systems, especially among top performers. This is despite the conformation of earlier meta-evaluations  which reported strong results for \umbrela{} on traditional IR \cite{thakur2025rag} (Tau@60=0.89), its effectiveness dropped sharply once systems included \llmjudge{} components (Tau@60=0.63 / Tau@10=0.38). These results highlight that unreliable agreement with human judges can incentivize misleading system design. Research on the robustness of such judges for RAG evaluation is limited.

\subsection{Nugget-based RAG Test Collections}

Many recent RAG benchmarks rely on nuggets that are produced by large language models.  Fully automatic approaches, such as the \rubric{}~\cite{farzi2024pencils}, generate nuggets without human oversight, while semi‑automatic pipelines, e.g., \AutoNuggetizer{}~\cite{pradeep2024autonuggetizer}, employ post‑hoc human filtering.  In both cases the nuggets remain traceable to an LLM source, which means a competent RAG system can simply reproduce or approximate them.  Consequently, showing alignment between system‑generated nuggets and gold nuggets yields little insight into the distortions introduced by nugget prediction.  

To properly interrogate RQ3, we thus require a test collection whose nuggets were 
 created without LLM influence; only then can we simulate the effects of nugget 
 prediction and measure the distortion it introduces into evaluation outcomes.
Collections that satisfy this criterion are scarce.  
Given this landscape, we turn to the TREC NeuCLIR 2024 Report Generation Pilot, to date, the only publicly available RAG benchmark that offers a fully manual nugget bank created without LLM assistance.  Details of this collection are deferred to the evaluation section.


\section{\crucible{}: A Subversion Probe for \autoargue{}}
\label{sec:crucible}

\begin{figure}[t]
\includegraphics[width=1\textwidth]{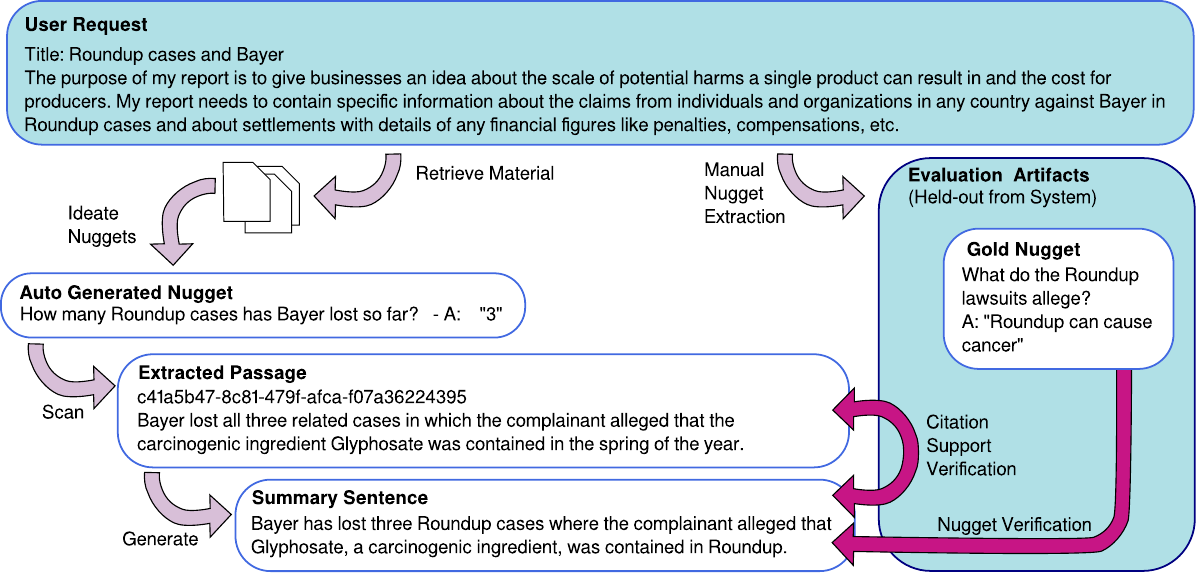}

\caption{Workflow of \crucible{} and its evaluation in the \autoargue{} system.
\crucible{} ideates nuggets from retrieved documents, extracts and summarizes passages, and assembles them into a cited report. 
\autoargue{} evaluates each sentence by checking coverage of manual gold nuggets and verifying citation support.
It is common that the summary sentence created to address a system nugget, such as 
``How many RoundUp cases has Bayer lost so far?'',
also covers gold nuggets for evaluation, such as 
``What do the RoundUp lawsuits allege?'', since both sets capture related aspects of the same topic.}

\label{fig:roundup-example}
\end{figure}

\subsection{\crucible{}: A RAGE System} \label{sec:crucible-rage}

As an experimental proofing ground, we design the RAGE system \crucible{} \cite{dietz2026incorporating}, a retrieval-augmented generation (RAG) system that incorporates ideas from automatic evaluation (E). 
\crucible{} adopts ideas from nugget-based evaluation, to use document-grounded nuggets to generate reports, as depicted in Figure~\ref{fig:roundup-example}. Hence, it is designed to optimize \emph{nugget-based} evaluation measures while providing highly faithful \emph{citations}---two key metrics measured by the evaluation system \ARGUE{}  (Section \ref{subsec:evaluation-measures}). 

\begin{enumerate}
    \item \textbf{Nugget Ideation.} \crucible{} starts with automatically generating a system nugget bank
    that drives the extraction and generation processes.
    In our setup, nuggets take the form of Q\&A pairs that represent topically relevant key facts. We purposefully imitate the style of Q\&A nugget used by \ARGUE{}.
    \item \textbf{Retrieval.} 
    Documents are collected from top 20 of PLAID-X dense retrieval mod\-el~\cite{yang2024translate}, a retrieval system provided by the organizers of TREC NeuCLIR, whose dataset we use in this study.
    We also study the effects of other retrieval models~\cite{nguyen2025milco,zhang2025qwen3},
    but these results are omitted for brevity.
    \item \textbf{Sentence Extraction.} \crucible{} extracts candidate passages from retrieved documents along with self-contained sentences
    using prompt-based templates\footnote{The system message was omitted due to a bug. Updated results in online appendix.} that align document content to nugget questions and answers.
    \item \textbf{Selection and Assembly.} Depending on the requested report length, \crucible{} selects up to $k\in\{1,5,20\}$ sentence(s) per system nugget
    with the highest extraction confidence
    and assembles them into a report that maximizes coverage of the system nugget bank,
    using the the source document of each sentence as a citation.
\end{enumerate}

LLM prompts provided in the online appendix and in the system paper \cite{dietz2026incorporating}.

\subsection{Worked Example}

Figure \ref{fig:roundup-example} provides an example of how \crucible{} generates a report for a user request about lawsuits and settlements. 
System-generated nuggets guide the extraction of matching passages, which are summarized and cited. A list of system and gold nuggets for this example is available in the online appendix.


The \ARGUE{} framework separately evaluates each included sentence along with its citation. The evaluation makes use of a test collection with manual gold nuggets, which are developed by human judges to determine the relevance of the information. An example is ``What do the RoundUp lawsuits allege?''. The \autoargue{} implementation uses an LLM prompt to determine whether a gold nugget is covered by the sentence and the citation.  Additionally, \autoargue{} verifies that cited documents support the corresponding summary sentence with a separate prompt. 

Figure \ref{fig:roundup-example}  depicts how the \crucible{} system generates one sentence for its response:  ``Bayer has lost three Roundup cases where the complainant alleged that Glyphosate, a carcinogenic ingredient, was contained in Roundup.'' This sentence was generated to address the system-generated nugget 
``How many RoundUp cases has Bayer lost so far?'', but because the document contains additional relevant information,
the resulting sentence also addresses the gold nugget from the test collection 
``What do the RoundUp lawsuits allege?''. 
This effect is expected, as both sets of nuggets represent concrete information elements for the same broad topic. Often there are closely related questions in both nugget sets, such as the number of lawsuits.

\begin{figure}[t]
\includegraphics[width=1\textwidth]{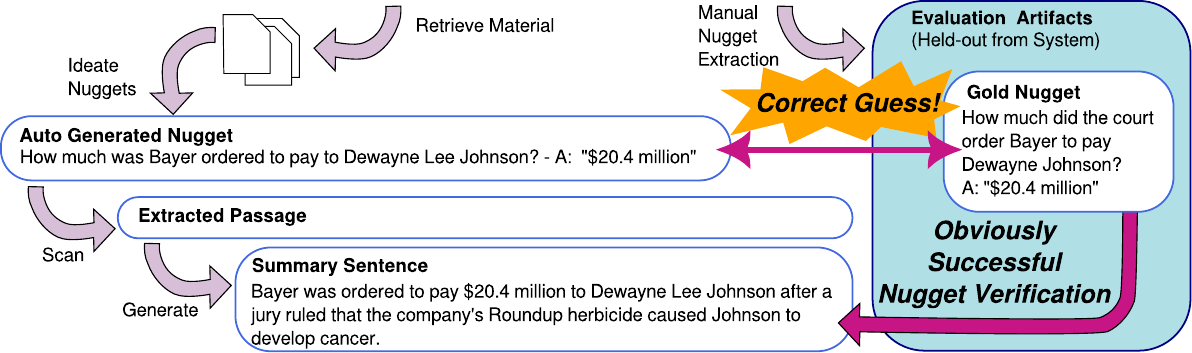}
\caption{Example where \crucible{} correctly guesses a gold nugget 
(``How much did the court order Bayer to pay Dewayne Johnson?''). 
RQ3 explores how evaluation outcomes change if RAG systems reliably guess the gold nugget set}
\label{fig:roundup-guess}

\end{figure}

In some cases, \crucible{} directly guesses a gold nugget, as shown in Figure \ref{fig:roundup-guess}. 
Although our nugget generation was not tuned to imitate gold nuggets, some overlap arises naturally. 
As semi-automatic pipelines increasingly use LLMs to propose nuggets for human verification, we expect such overlap to be increasingly common. 
To study the downstream effect in RQ3, we simulate a setting where \crucible{} perfectly predicts all gold nuggets and observe the effects on \autoargue{} evaluation scores.


\subsection{\crucible{} \TestProbes{}}

To ``attack'' potential vulnerabilities of \ARGUE{}, we design subversion probes that augment the \crucible{} system. This work's key innovation is the use of insider knowledge about the LLM judge's
(here: \autoargue{}'s) own decision process to filter and rewrite outputs,
ensuring that every retained sentence is ``approved'' by that judge. 
Our hypothesis is that this leads to measurable effectiveness increases
that may or may not generalize to improvements under human judgment.

We design two kinds of \testprobes{} to simulate what would happen \emph{if} knowledge of specific aspects of the \autoargue{} \llmjudge{} became known to system developers.

\subsubsection{Guessed prompt \probe{}.}
The first \testprobe{} simulates the condition in which the prompts used by the judge are becoming public knowledge.
Such prompts could then be used to improve the quality of the responses from the perspective of the \llmjudge{} prior to evaluation.
This \probe{} applies two filtering steps directly after the \textbf{Sentence Extraction} step (Section \ref{sec:crucible-rage}):
\begin{enumerate}
    \item \textbf{Citation Filtering.} Each candidate sentence and extracted passage gets passed through \autoargue{}’s  own citation support checks. 
    \item \textbf{Nugget Coverage Filtering.} Each candidate sentence (or extracted passage) is passed through \autoargue{}’s nugget matching checks, to verify that it covers the intended system nugget.
\end{enumerate}
Candidate sentences that fail these checks are discarded before \textbf{Selection and Assembly}. We use these \probes{} to study the change in evaluation measurements from using the exact prompts of the \llmjudge{} (RQ2).

\subsubsection{Guessed nuggets \probe{}.}
The second \testprobe{} simulates the condition in which gold nuggets are guessable by a fully automatic system (RQ3).
This could arise when relevance for test topics is well understood by LLMs
or gold nuggets are taken from a pool of automatically predicted system nuggets.
This could arise in a variety of ways---e.g., via leaked knowledge of the model used to generate the nuggets or of other aspects of the nugget curation process. Here, we consider the limiting case of such a leak, in which the \emph{exact} set of gold nuggets is known. 
To simulate this condition, this \probe{} replaces the \textbf{Nugget Ideation} step (Section \ref{sec:crucible-rage}) with the set of gold nuggets. This point highlights the importance of keeping gold nuggets secret.

\subsection{What is the point of \testprobes?}

\paragraph{System.}
From the system perspective, \crucible{}'s \testprobes{} act as a wrapper;
any RAG pipeline can be run through them to produce outputs tuned for \autoargue{}.
By design, the wrapped systems obtain higher evaluation scores under the \llmjudge{}.
Whenever these materialize as genuine improvements under human judgment, we suggest that the \probe{} is incorporated into the wrapped system to yield an improved RAGE system.



\paragraph{Evaluation.}
From the evaluation perspective, \crucible{} operationalizes an ``attack’’;
if a system that is adulterated with \crucible{}'s guessed probes achieves much higher \argue{} scores but not higher manual evaluation scores,
those improvements reveal a vulnerability in the attacked evaluation method.
In such cases, guardrails to thwart the attack should be developed,
such as ensembles of prompts, hidden nugget banks, or held-out LLMs.
The \probe{} would then be used to test the resilience of the evaluation method and the efficacy of the proposed guardrails.

\paragraph{\Metaeval{}.}
Paradigm changes in system development present a challenge to the reliable design of \metaeval{}s for \llmjudge{}s.
In such cases, high agreement between an \llmjudge{} and human evaluation can be observed
only for certain systems, e.g., those that did not adopt ideas from LLM judge approaches.
However, as demonstrated by \citet{clarke2024llm}, this correlation may not hold for next-generation systems
for which we seek an \llmjudge{} that can reliably identify progress.
Especially when the \llmjudge{} quality is measured via Kendall's tau correlation between automatic and manual leaderboards, a healthy variety of approaches is essential. Here \testprobes{}  that can modify the outputs of a set of existing systems provide a force-multiplier that can simulate anticipated paradigm-changes of RAG systems, rendering the \metaeval{} more convincing.



\section{Experiment Setup}


\subsection{Dataset}
We use the test collection of the TREC NeuCLIR 2024 report‑generation pilot collection~\cite{lawrie2024overview}.  The corpus contains ten million documents in Chinese, Russian, and Farsi, each accompanied by an automatic English translation.  For each of the 19 report requests (queries), assessors provided a title, a problem statement, and the requester's background.  

For all report requests, TREC assessors manually designed a comprehensive set of nuggets,
each comprising a factual question and a set of correct answers.
The nuggets were created after assessors read the relevant documents in the original language and \emph{did not involve any LLM assistance}.  
On average, 10–20 question‑answer pairs were produced per request, and these nuggets serve as the gold standard for assessing system‑generated reports submitted to the 2024 NeuCLIR track (our system's reports were not part of the original pool).  
We focus exclusively on this dataset because, at the time of writing, no alternative RAG benchmark provides a comparable manually curated nugget bank.  

\subsection{Evaluation Measures}
\label{subsec:evaluation-measures}

The \autoargue{} evaluation system used manually designed gold nuggets from the TREC NeuCLIR track
to determine the relevance of the system's generated report.
Using \texttt{Llama3.3-70B-Instruct}, \autoargue{} scans each sentence of the report for mentions of a correct answer to the nugget question.
The nugget recall metric follows the assumption that the more nuggets are covered in the generated report,
the more relevant the report is.  
Nugget-measures are complemented by a verification of citation faithfulness with the LLM. 

We report results using \autoargue{} for the following measures:

\begin{description}
    \item[Nugget Recall] $=\frac{\text{covered nuggets}}{\text{gold nuggets}}$ measures the recall of gold nuggets.
    \item[Nugget Density] $=\frac{\text{covered nuggets}}{\text{sentences}}$ measures the tradeoff between concise summaries and nuggets covered. 
    \item[Relevant Sentences] $=\frac{\text{sentence with nuggets}}{\text{all sentences}}$ measures how many sentences include facts relevant for the report.
    \item[Citation Support] $=\frac{\text{supported citations}}{\text{citations}}$ of citations supporting their sentence.
\end{description}

\subsection{Variations and Baseline Systems}

To explore our hypotheses, we compare several \crucible{} variants; those marked with $\dagger$ rely on insider knowledge about evaluator prompts (RQ2);
those marked with $\star$ rely on insider knowledge to generate nuggets (RQ3):

\begin{description}
    \item[Base:] Basic \crucible{} system using PLAID-X~\cite{yang2024translate} as the retrieval model,
    no sentence filtering, and automatically generated system nuggets.\footnote{We also experimented with multilingual LSR~\cite{nguyen2025milco} and Qwen3~\cite{zhang2025qwen3} and found similar trends. Results are omitted here, but available in \citeauthor{dietz2026incorporating} \cite{dietz2026incorporating}.}
    \item[Citation Filter~$\dagger$:] Filter citations that don't support the sentence using \autoargue{} prompts (and removing sentences lacking any citation).
    \item[Cov-Sentence~$\dagger$:] Filter sentences that do not cover any of the system nuggets.
    \item[Cov-Extract~$\dagger$:] Filter sentences where the extracted passage does not cover any of the system nuggets.
    \item[Gold nuggets~$\star$:] Simulate the system correctly guessing all gold nuggets.   
\end{description}

For each variation, we compare reports of different lengths:
\begin{description}
    \item[Short:] Up to $k=1$ sentence per nugget (report limit=2000 characters)
    \item[Medium:] Up to $k=5$ sentences per nugget (report limit=10,000 characters)
    \item[Long:] Up to $k=20$ sentences per nugget (report limit=1,000,000 characters)
\end{description}

To provide a reference point, we compare our \crucible{} system variations to the following systems:

\begin{description}
    \item [GINGER \cite{lajewska2025ginger}:] A nugget-based RAG system using GPT-4o with clustering. Designed for TREC RAG 2024 topics and evaluation with \AutoNuggetizer{}~\cite{pradeep2024autonuggetizer}. Using 100 documents retrieved by Qwen3 (best tested variant).
    \item [GINGER-Llama:] \ginger{} using the same LLM
          as \crucible{} (\llama{}).

%
    \item[GptResearcher \cite{duh2025hltcoe-liverag}:] A simple RAG pipeline implemented
    using the open source \gptresearcher{}~\cite{Elovic_gpt-researcher_2023} toolkit
    that simply passes the top-ranked retrieved documents to \llama{} to generate the final report. 
%
    \item[BulletPoints~\cite{yang2025hltcoetrec}:] An extractive pipeline (submitted as \texttt{hltcoe‑eugene}) that was one of the top‑performing systems in the TREC NeuCLIR 2024 Report Generation Pilot.
    The {\sc BulletPoints} pipeline extracts and groups facts from the retrieved documents using \llama{}. 
\end{description}

We use \llamabig{} for all system implementations and the \autoargue{} evaluation, except where otherwise noted.

\section{Results and Analysis}



\subsection{RQ1: Does knowing the evaluation framework and metric improve the measured performance? (Yes)}

\begin{figure}[t]
    \centering
    \includegraphics[width=0.49\linewidth]{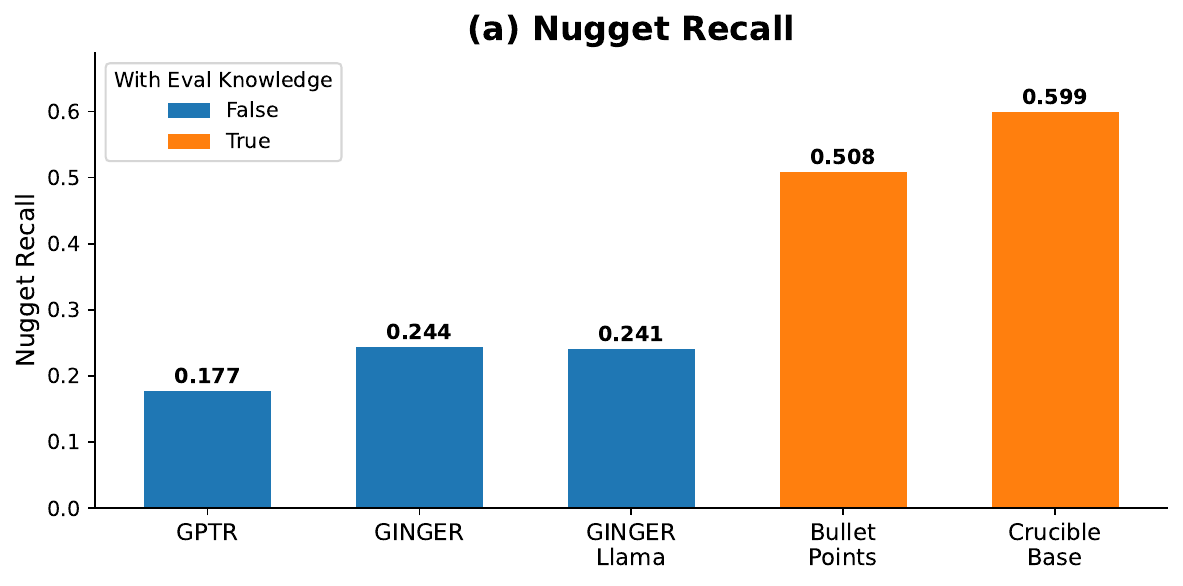} \includegraphics[width=0.49\linewidth]{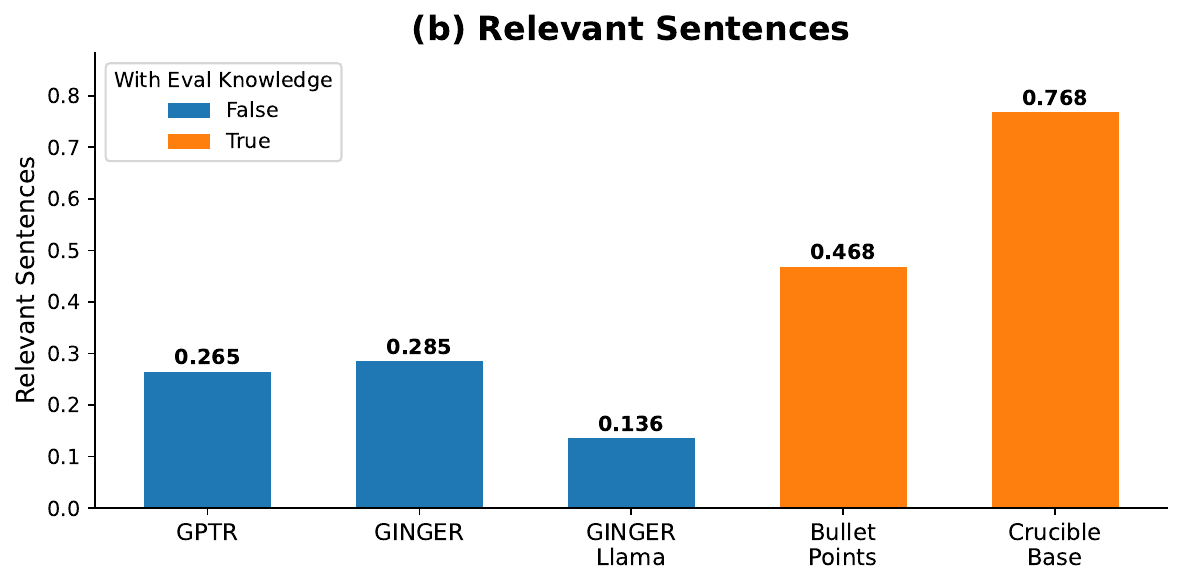}

    \includegraphics[width=0.49\linewidth]{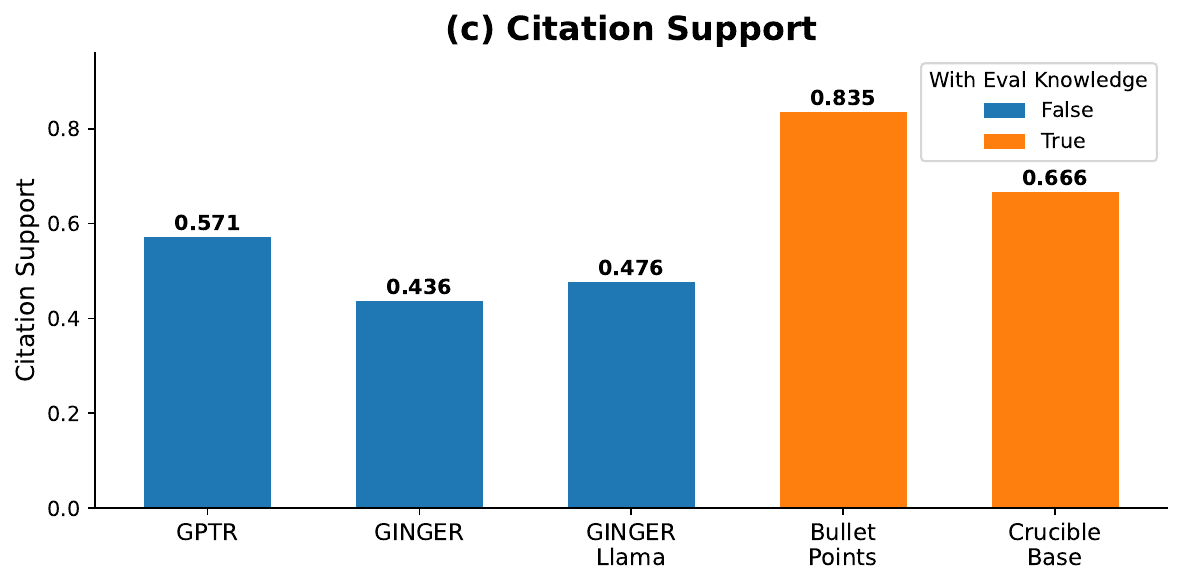} \includegraphics[width=0.49\linewidth]{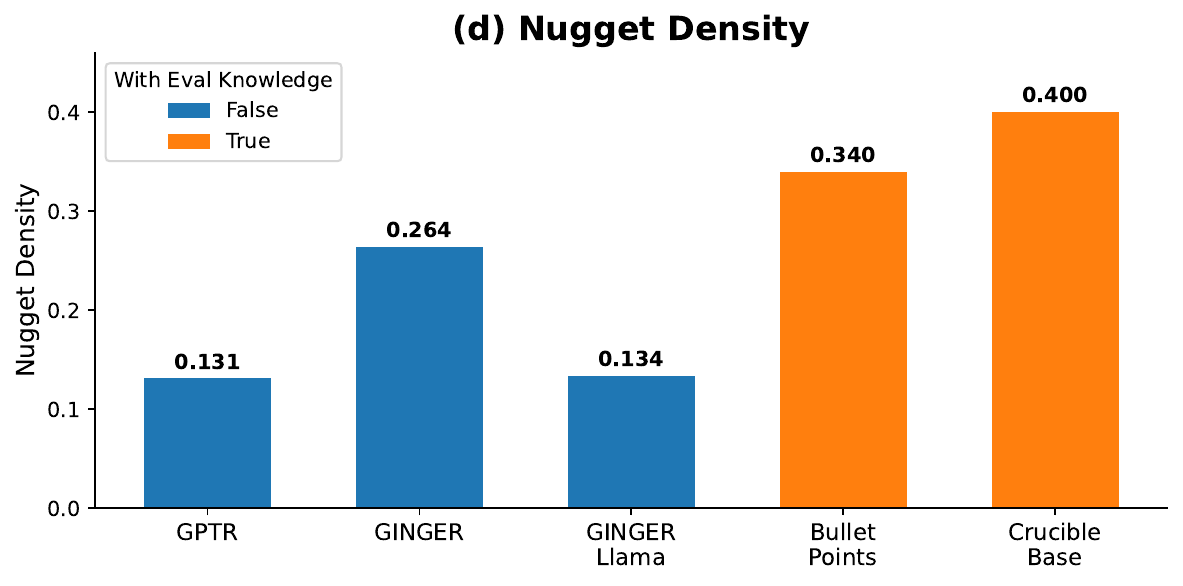}
    \caption{RQ1: Yes, knowledge of the evaluation system is likely to help development of a RAG system that obtains high evaluation scores.
    \ginger{}, a recent nugget-first RAG system designed for TREC RAG 24 is expected not to have used any insider knowledge about the \argue{} evaluation framework. We show that despite conceptual similarities between \ginger{} and \crucible{}, the performance characteristics are vastly different.}
    \label{fig:rq1}
\end{figure}

We explore RQ1 by comparing RAG systems designed for the TREC NeuCLIR task and \autoargue{}-based evaluation
(``with evaluation knowledge'')
to RAG systems designed for a different task. The latter set of systems includes both \gptresearcher{} and \ginger{},
which was designed for the TREC RAG 24 task that was evaluated with the AutoNuggetizer system~\cite{pradeep2024autonuggetizer}.
We also include {\sc BulletPoints}, a top performing system from the NeuCLIR track.

Figure \ref{fig:rq1} presents these systems in order from no to full knowledge  of the \argue{} evaluation approach. The more general system \gptresearcher{} generally lags behind.
 Even for the two nugget-first RAG systems \ginger{} (no eval knowledge) and \crucible{} (with eval knowledge), the fact that evaluation knowledge is helping is clearly visible in relative performance gains ranging from at least +16\% to +100\% across all metrics. Thus, system designers who optimize for an evaluation framework will likely see measurable improvements.


\subsection{RQ2: Does knowing the prompt used in the automatic evaluation improve measured performance? (Yes, weak signal)}



\begin{figure}[t]
    \centering
    \includegraphics[width=0.49\linewidth]{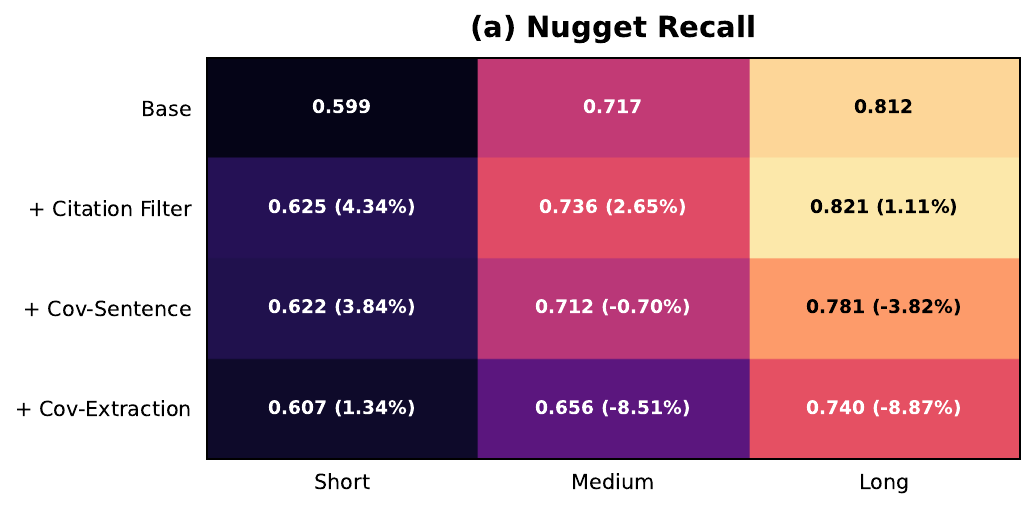} \includegraphics[width=0.49\linewidth]{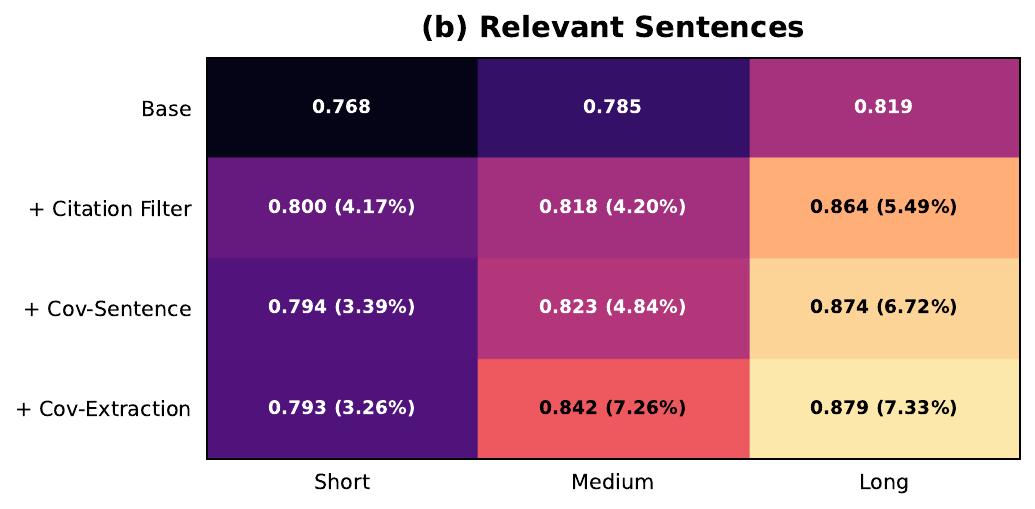}

    \includegraphics[width=0.49\linewidth]{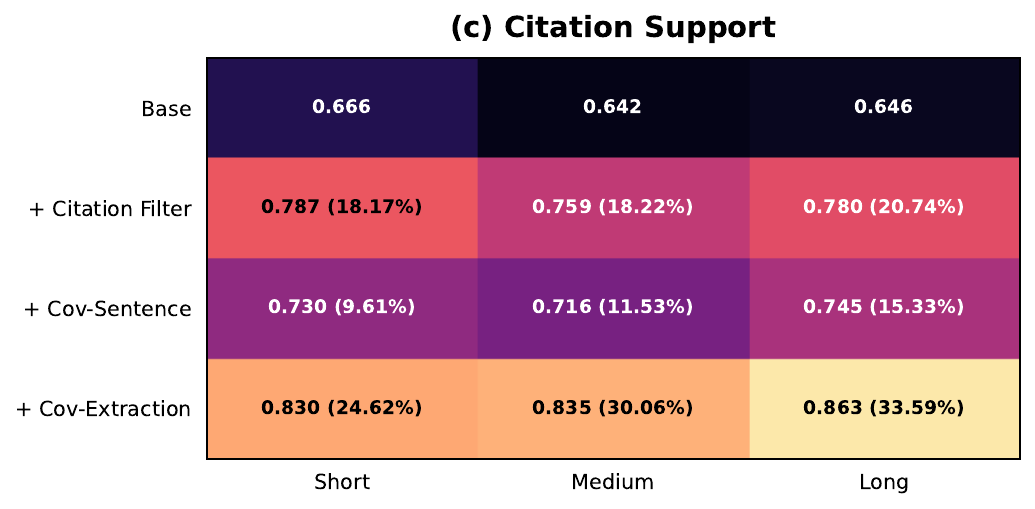} \includegraphics[width=0.49\linewidth]{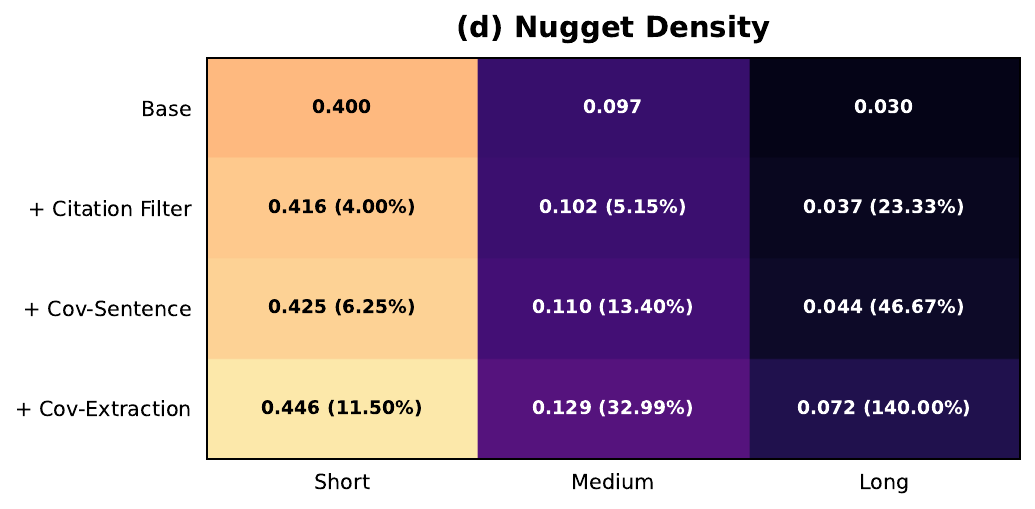}
    \caption{RQ2: Yes, with the exception of nugget recall, filtering candidate extractions with the evaluation prompt for citation support and/or nugget detection consistently improved the respective evaluation metrics. \% indicates relative improvement over ``Base''. 
    }
    \label{fig:rq2}
\end{figure}

\begin{figure}[t]
    \centering
    \includegraphics[width=0.49\linewidth]{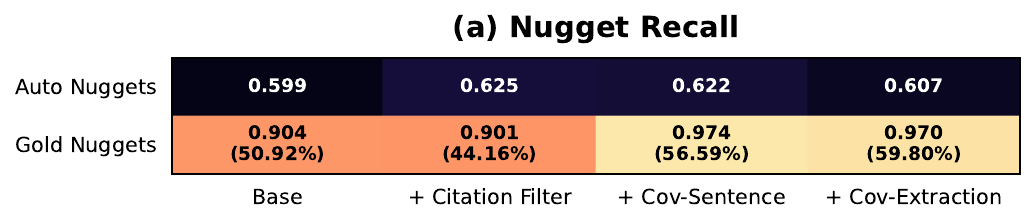} \includegraphics[width=0.49\linewidth]{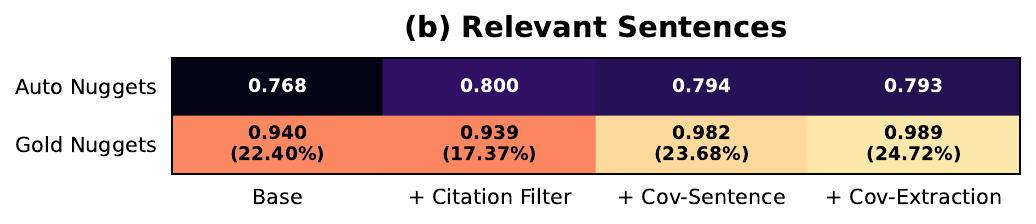}

    \includegraphics[width=0.49\linewidth]{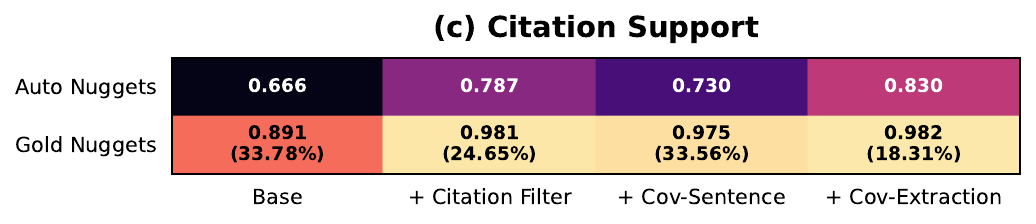} \includegraphics[width=0.49\linewidth]{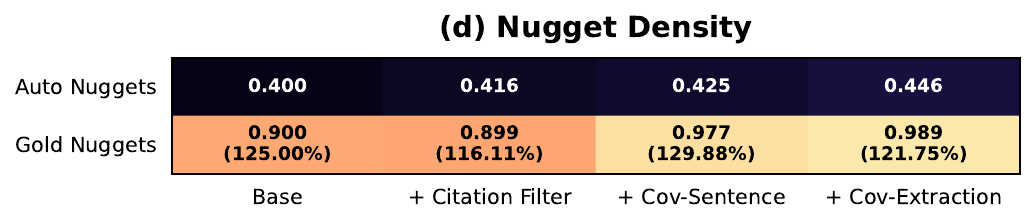}
    \caption{RQ3: Yes, a system that could predict gold nuggets, would obtain a stark increase across all metrics and, when combined with insider knowledge of evaluation prompts, achieve near-perfect scores of $\approx 0.99\%$ in all metrics. These results are likely tainted by circularity.}
    \label{fig:rq3}
\end{figure}


For RQ2, we ask whether the prompt-guessing probes yield the expected improvement in \autoargue{}'s evaluation measures. 
While the \crucible{} base system uses its own prompts for the initial sentence extraction,
we evaluate whether filtering these sentences with \autoargue{}'s prompts leads to higher evaluation scores, which could indicate a circularity.
We explore this question in several variations of the \crucible{} system,
generating short,
medium
and long reports.

We verify that the citation filtering probe indeed leads to an +18\%-+20\% improvement under \autoargue{}'s citation support metric (Figure  \ref{fig:rq2}c). Interestingly, removing sentences where the cited document does cover the same nuggets as the sentence (+Cov-Extraction) further improves citation support to +24\% to +33\%.

 The nugget cover filtering probe results in improvements under nugget density (+11\% / +33\% / +140\%)  and sentence relevance (as measured by nugget coverage) with gains between +3\% and +7\% (Figures \ref{fig:rq2}b, \ref{fig:rq2}d). Since our filter is only removing sentences, it is expected to see a decrease in nugget recall (Figure \ref{fig:rq2}a), even thought this can help for short reports.



Across all variations we observe consistent performance improvements, many of which are
confirmed by paired t-test.
We remark that by its bottom-up design,
the \crucible{} base system is already very effective
in grounding sentence generation in cited sources and system nuggets.

\subsection{RQ3: Does the ability to predict manually created gold nuggets
    improve the measured performance? (Yes, for nuggets)}

For RQ3 we explore a hypothetical scenario where the RAG system is able to predict
(or otherwise obtain)
the set of gold nuggets used by the evaluator.
We simulate this by using the gold nuggets as system nuggets in \crucible{}.

Figure~\ref{fig:rq3} demonstrates large improvements  (+44\% to +59\%) on nugget recall when the evaluator's gold nuggets are known. Especially when combined with the prompt filtering, ``perfect'' nugget recall, nugget density, sentence relevance, and citation support scores of 0.97--0.99 can be obtained (!).

\subsection{Are Improvements Genuine?}

We remark, that we do not believe that these results represent real achievements---these evaluation results are artifacts exploiting insider knowledge of evaluation secrets.

We note that the detection of nuggets in text is affected by the randomness of the underlying LLM, which can lead to disagreement between system extraction and \autoargue{} evaluator.
Manual inspection showed that the sentences covered both system and gold nuggets.

\ld{we should re-do this!}
Although all of our quantitative results are based on manual nuggets,
we are curious whether measured evaluation improvements would translate to a human-felt quality difference.
The author team selected two reports generated for the example Topic 335,
one from the \crucible{}-base variant
and one that applies both the guessed prompt and the guessed nugget probe.
The team is divided on which contains more relevant facts,
appreciating that the \crucible{}-base report mentions a wide range of numerical data on court cases,
both pending and lost at different court levels and dollar amounts of settlements.
However, the guessed prompt probe indeed translates to sentences
where extracted passages clearly specify the relevant entities, such as Bayer.
A team member commented that they considered the citations arising from this probe
as more faithful, and hence were less suspicious that information is incorrect---a clear indicator that future RAGE systems should adopt the Citation Filtering probe.




\subsection{Early Results from TREC 2025}

To gain empirical insights from fresh test collections, we submitted several variations of our approach to TREC 2025
\citep{dietz2025hltcoe}. 
In TREC RAGTIME, we find that investing effort in improving the system nuggets yields higher nugget recall ($0.27 \rightarrow 0.33^\blacktriangle$). In TREC DRAGUN, nugget coverage filtering results in small but not statistically significant gains in support ($0.166 \rightarrow 0.173$).

For TREC RAGTIME, citation filtering significantly improves citation support
($0.83 \rightarrow 0.99^\blacktriangle$), with no difference across approaches ($0.97\text{--}0.99$). 
In contrast, manual citation support annotations in TREC RAG show a substantial spread across methods, with weighted precision ranging from $0.59\rightarrow 0.69 \rightarrow 0.80^\blacktriangle$.  We remark that citation filtering introduces circularity with \autoargue{}'s evaluation procedure used in TREC RAGTIME. This divergence 
implies that our attack successfully exposes a prompt-based vulnerability.

\section{Conclusion}

We present a systematic empirical study of evaluation subversion using insider knowledge, not just a system description. \crucible{} is a RAGE system with a subversion probe
developed for the TREC NeuCLIR track and its evaluation framework \autoargue{}.
Using \crucible{}, we examine how insider knowledge about the evaluation process can distort measured performance. Specifically, we study three types of insider knowledge:
(RQ1) structural understanding of the evaluation targets;
(RQ2) knowledge of evaluation prompts;
(RQ3) knowledge of gold nuggets used for evaluation.
All three influence evaluation outcomes, with knowledge of the gold nuggets posing the most severe threat to validity.

While this work represents only an initial step toward a systematic investigation of vulnerabilities
in LLM-based evaluation paradigms such as \autoargue{}, 
it raises an important warning: the public release of evaluation artifacts,
including prompts and nuggets, carries significant risk. 
If evaluation artifacts are becoming public knowledge or are reverse-engineered during system development,
they can artificially inflate scores of systems developed with insider knowledge, undermining the integrity of empirical comparisons across systems.
Our study underscores the need for safeguards to prevent such subversion,
lest evaluation results become misleading or invalid.

To mitigate the vulnerabilities identified in this work while preserving re\-usable, transparent, and reproducible RAG evaluation, we recommend sharing test collections through platforms that support blinded experimental evaluation. The TREC Auto-Judge track provides one such example by using TIRA \cite{froebe2023tira} and by distributing our test \probes{} within this blinded evaluation setting.
We hope that future work will develop subversion probes for additional \llmjudge{} paradigms to quantify evaluation vulnerabilities more systematically and to inform the design of effective safeguards.

\paragraph{Disclosure of Interests.} Authors have no competing interests.

\bibliographystyle{splncs04nat}
\bibliography{bibio}

\end{document}